\documentclass[adp,fleqn]{w-art} 
\usepackage{times}
\usepackage{cite}
\usepackage{w-thm}

\usepackage[]{graphicx}
\chardef\bslash=`\\ 

\newcommand{\secref}[1]{Section~\ref{#1}}

\newcommand{\tpr}{transport process} 
\newcommand{\tprplur}{\tpr{}es}

\usepackage{float}
\floatstyle{ruled}
\newfloat{textbox}{tbhp}{bxl}
\floatname{textbox}{Box}

\begin{document}
\renewcommand{\copyrightyear}{2009}
\DOIsuffix{theDOIsuffix}
\Volume{12}
\Issue{1}
\Copyrightissue{01}
\Month{01}
\Year{2009}
\pagespan{1}{}
\Receiveddate{XX September 2009}
\Accepteddate{XX XXX 2009}


\title[Network modules help the identification of signaling pathways]{Network modules help the identification of key transport routes, signaling pathways in cellular and other networks}


\address[\inst{1}]{Semmelweis University, Department of Medical Chemistry, Budapest, Hungary}

\author[R.\ Palotai]{Robin Palotai\footnote{E-mail: {\sf palotai.robin@gmail.com}}\inst{1}}
\author[P. Csermely]{Peter Csermely\footnote{Corresponding
     author \quad E-mail: {\sf csermely@eok.sote.hu}, Phone: +36\,1\,459\,1500\,/\,60130
     Fax: +36\,1\,266\,2650}\inst{1}} 

\begin{abstract}
Complex systems are successfully reduced to interacting elements via the network concept.
Transport plays a key role in the survival of networks --- for example the specialized signaling cascades of cellular networks filter noise and efficiently adapt the network structure to new stimuli.
However, our general understanding of transport mechanisms and signaling pathways in complex systems is yet limited.
Here we summarize the key network structures involved in transport, list the solutions available to overloaded systems for relaxing their load and outline a possible method for the computational determination of signaling pathways.
We highlight that in addition to hubs, bridges and the network skeleton, the overlapping modular structure is also essential in network transport.
Moreover, by locating network elements in the space of overlapping network modules and evaluating their distance in this `module space', it may be possible to
approximate signaling pathways computationally, which, in turn could serve the identification of signaling pathways of complex systems.
Our model may be applicable in a wide range of fields including traffic control or drug design.
\end{abstract}
\maketitle

\section{Introduction}
\label{intro}

The network concept is successfully applied to reduce complex systems into a set of interacting elements connected by links to examine, understand and predict the topology, dynamics and emergent properties of the system\cite{evolnet-dorogovtsev02,cellfunc-barabasi04,boccaletti-2006,csermely-06}. In most networks the elements are autonomous agents, which not only develop direct interactions via links, but also establish long-range indirect interactions via various \tprplur{}.
The necessity of a \tpr{} is usually evoked by a need for communication (social and telecommunication networks\cite{orginfo-radner93,scott-social-network-analysis}), transfer of resources (logistic networks, power grids\cite{fastroad-schutes07,Watts:1998}) or regulation ensuring a fast, magnified and efficient response (signal transduction networks\cite{bionet-zhu07,bionet-almaas07}).
The \tpr{} is not only an emergent property of the network, but also a significant organizing force behind the structure and dynamics of the network. Links of the network may emerge in order to serve the \tpr{}\cite{optimal-structures-cabrales-2002} and disappear later, if their existence is not required anymore or becomes even harmful\cite{cascade-control-motter-2004}.

Network signaling may be considered as a highly specialized case of transport. Signaling of cellular networks is a system level response to an incoming stimulus and is an extremely selective behavior fine-tuned by evolution. It efficiently filters noise-like stimuli, while quickly develops complex signaling cascades in response to a recognized stimulus\cite{genanal-gerstein06,szalay-2007-581}.

How does a network learn to discriminate between signal and noise? Our own studies\cite{szalay-2007-581,palotai-2007,mihalik-2008,csermely-perturb-2009} may help us to describe a common scenario: when an unusual signal arrives, which is strong and persistent enough to modify network behavior, the network slightly or profoundly disassembles: as a major process network modules (groups, communities\cite{boccaletti-2006,palla-2005-435}) become loosely attached with a decreased overlap. When the stimulus is over, the network reassembles again. In this phase a large number of inter-modular contacts become re-established. However, these inter-modular contacts will not be exactly the same as before the stimulus: by developing a structural `imprint' of the signal, the complex system has now a memory, it learned, on one hand which links may be more effective to dissipate the stimulus most efficiently, and on the other hand, which links are disturbing this process. If a similar stimulus arrives regularly (or the stimulus is large enough that all networks which were unable to learn the reorganization described above will disassemble and die) than the newly selected pathway may become dominant and may behave as a signaling pathway from then on.

In~\secref{sec:char_transp} we enumerate the main properties differentiating \tprplur{} and consider optimality criteria with an emphasis on network throughput,
then summarize the basic structures utilized by the network for efficient transport and filtering, namely the network skeleton, hubs (highly connected elements),
bridges (elements connecting sparsly inter-connected network segments) and network modules (communities).
In~\secref{sec:sim_pathway} we propose a method for reconstructing simulated and signaling pathways based on overlapping network module information.
In~\secref{sec:summary} we summarize our findings and conclude.

\section{Characterization of network transport}
\label{sec:char_transp}

\begin{vchtable}
\vchcaption{Main properites of \tprplur{}}
\label{tab:tpr_prop}\renewcommand{\arraystretch}{1.5}
\begin{tabular}{p{0.25\textwidth} p{0.65\textwidth}} \hline
\textbf{Property} & \textbf{Description} \\
\hline
Purpose & Goal of the \tpr{}, usually related to the fitness or survival of network elements and the network as an entity. \\
Sources and sinks & Specific network elements may be identified as the source or sink of a given transported quantity. \\
Information need & Routing mechanism of network elements determining which neighboring element will they forward a received quantity to 
				may require either local, global or intermediate~(mesoscopic) knowledge about the network. \\
Determinism & Routing may be deterministic or stochastic. \\
Adaptiveness & Routing may~(adaptive process) or may not~(static process) be affected by the dynamic properties of the network. \\
Information preservation & Quantities transported may remain unchanged, suffer distortion or even get lost. \\
Time & Transport may be a discrete- or continous-time process. \\
\hline
\end{tabular}
\end{vchtable}

As described in Table~\ref{tab:tpr_prop} summarizing the main properties differentiating \tprplur{},
the goal of the \tpr{} is usually related to the survival of the system defining the network and therefore transport is related to the survival of the network as a connected graph with a large giant component.
Transport may mobilize resources or information between network elements which are then used for the benefit of the system described by the network. For example, a network element in need may propagate a request message and other elements may send resources in response --- this signaling scenario could emerge without network elements having attributable intentions or desires.
`Transport-provoking' cooperation may emerge through evolutionary mechanisms like signaling games\cite{zahavi-75}. The recently introduced protein games might also play a similar role in case of amino-acid networks\cite{kovacs-2005,csermely-perturb-2009}.

As for routing, local mechanisms tend to be stochastic because they lack extended information and therefore cannot be certain about the effectiveness of any single deterministic choice, while informed global routing mechanisms usually favor more deterministic approaches.
Cellular networks exhibit local routing property, as the transmitted signal can be represented by the propagation of conformational changes  of interacting proteins,
and such changes may be evaluated locally via means of induced fit, conformational selection or protein games\cite{inducedfit-koshland-58,conf-selection-75,csermely-perturb-2009}.

\begin{vchtable}
\vchcaption{Expectations toward \tprplur{}}
\label{tab:tpr_expected}\renewcommand{\arraystretch}{1.5}
\begin{tabular}{p{0.15\textwidth} p{0.6\textwidth}} \hline
\textbf{Expectation} & \textbf{Description} \\
\hline
Soundness & Operation of the \tpr{} should strive to achieve and maintain its goal. \\
Simplicity & Among \tprplur{} of similar performance the one with the simplest mechanism is preferred. \\
Robustness & The \tpr{} should resist network failures, or at least degrade gracefully, proportional to network load or damage. \\
Stability & Operation of the \tpr{} should lead to an equilibrium state of the network under stable circumstances. \\
Fairness & The \tpr{} should strive to satisfy the transportation needs of all network elements equally. \\
Optimality & Operation of the \tpr{} should be optimal for a set of criteria. \\
\hline
\end{tabular}
\end{vchtable}

Let us take a detour and examine the optimality criterion of Table~\ref{tab:tpr_expected}, which lists common expectations originally set towards computer networks\cite{tbaum}.
It should be noted that an universally applicable set of optimality criteria does not exist, mainly due to that different optimality criteria are usually in conflict with each other. Conventional optimality criteria include low duration of delivery, short delivery paths or high transport throughput.
In Box~\ref{transport_model_box} we summarize a simple, yet descriptive model of network transport, which is sufficiently abstract not to distract attention with implementation details, but still lets us draw conclusions about the network transport processes focusing on the criteria mentioned above\cite{wang-survey-2007,optimal-structures-cabrales-2002,effrouting-yan-2006,tadic-2006}.

For example, in cell signaling networks short delivery paths would be preferred to reduce the distortion or loss of information (and indeed, most cellular networks are small-worlds\cite{Watts:1998,cellfunc-barabasi04,csermely-06}), while higher throughput would let the signaling network handle more stimuli simultaneously. Unfortunately shorter delivery paths increase the load on network elements of high centrality, and this, in turn, lowers the maximum possible throughput.
As the example described here shows a \tpr{} may conform to different optimization criteria to some extent but not all of them simultaneously. Moreover, if the satisfaction of multiple 
optimization criteria involves an increased complexity of the \tpr{}, this increased complexity may hurt our expectations of simplicity, robustness or stability.

Knowing that the topology of any network sets an implicit upper bound on the maximum possible network throughput\cite{bottlenecks-stanley-2006},
it is interesting to investigate what kind of measures could the network utilize --- apart from rearranging or coarsening its link structure\cite{cascade-control-motter-2004,respredict-motter-2008,silencing-motter-2008} --- in order to relax overloaded elements and prevent congestion.

First, network elements may exhibit adaptive behavior of taking into account the load of other elements in their routing mechanism.
This behavior is exemplified by the multiple copies of protein isoforms in critical positions of cellular networks,
such as the `critical nodes' defined by Kahn and co-workers\cite{critsignal-kahn-2006}.

Second, network elements may resist to transport more than a given quantity, resulting in a filtered, faulty transport process but relaxing the load on the elements of the network.
It is not surprising that network elements and structures of high centrality (having consequently a high load) are natural candidates for such filtering, because
these elements of high centrality are expected to constitute a network skeleton or superhighway of transport\cite{superhighway-wu-2006} and thus
are able to filter excessive amounts of transported quantities.
Generally, congestion affects most the communication boundaries, such as central hubs of hierarchical networks or overlaps of network modules, both providing bridges between different network segments\cite{instabilities-gfeller-2005}.
If we define modules as having more intra-modular links than inter-modular\cite{radicchi-2004-101}, then modules themselves also act as noise traps with noise rather circulating inside the module and eventually getting dissipated instead leaving the module.

Third, the routing mechanism may decide to sacrifice certain optimization criteria in favor of network throughput by deliberately utilizing
alternative or back-up routes to some extent in parallel with the network skeleton. This procedure is not necessary adaptive, for example
(overlapping) network module information may serve as a basis for static routing if known\cite{impact-community-danon-2008,msc-palotai-2008}.

\begin{figure}
\includegraphics[width=\linewidth]{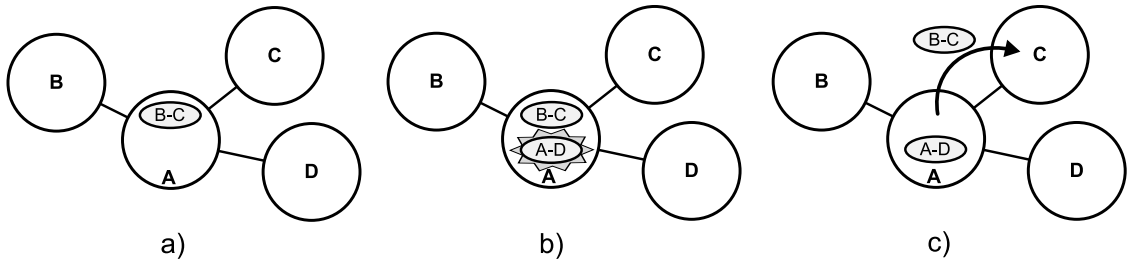}
\caption[Example for simple model of transport]{An example scenario for the simple model of Box~\ref{transport_model_box}.
\textbf{a)} The network element A and its neighboring elements B, C and D are shown. The packet pool of element A is shown containing a packet denoted B--C with source element~B and sink element~C.
\textbf{b)} Generation step: The new packet~A--D is generated at element~A with random sink element of~D.
\textbf{c)} Transport step: A random packet (now packet~B--C) is selected from the packet pool of~A and becomes forwarded according to the routing mechanism.  As packet~B--C reaches its sink element~C, it will disappear from the network.
}
\label{fig:1}
\end{figure}

\begin{textbox}[t]
\caption{\small A simple model of network transport}
\label{transport_model_box}
In the original model of\cite{optimal-structures-cabrales-2002} information packets denoted $a^{ik}$ are traveling from the source network element~$i$ to the sink network element~$k$ in the network $G=(N,E)$. In each discrete timestep first $R=\rho{} N$ new packets~$a^{ik}$ are generated with $i$~and~$k$~chosen randomly and are added to the pool of packets at~$i$ denoted~$Q_i$. Then for each element~$u$ a count of packets~$C_u=C$ are randomly removed from~$Q_u$ and forwarded according to the routing strategy, or to the sink~$k$ if it is neighboring~$u$ (and thus the packet is removed from the network). Fig.~\ref {fig:1} shows an example scenario.
\\ \par

This discrete transport model is very flexible: First, both sources and sinks can be identified, however this is not strictly necessary for its application. Second, depending on the applied routing strategy, the information need may be either local, global or mesoscopic, transport may either be stochastic or deterministic, information may be preserved or lost with some probability at each step of routing (see Table~\ref{tab:tpr_prop}). 
\\ \par

If the routing strategy is static and Markovian (the packet routing is independent of previously visited network elements), then both the expected path length and the expected load~$B_u$ of any element~$u$ (called \textit{effective betweenness}) can be analytically derived.
Moreover, if we define congestion as a state where exists an element~$u$ in the network with the packet pool~$Q_u$ growing faster than the processing capacity~$C_u$ of that element, then the throughput of the transport process can be characterized with the maximum $R=R_c$ value without congestion, given by $R_c=\min{\lbrace{}C_u N (N-1) / B_u\rbrace{}}$. Note that if any~$C_u=C$ then the network throughput is capped by the element of maximum effective betweenness.

\end{textbox}

\section{Simulated pathways on overlapping modules}
\label{sec:sim_pathway}

Recent advances in network module identification methods are not only able to assign elements to multiple overlapping modules but also provide metrics describing membership strength of any elements to different modules\cite{palla-2005-435,reichardt-2004-93,kovacs-2006-pattern,nepusz-2007,zhang-2007},
effectively locating network elements in the $M$-dimensional module space, $M$~being the number of modules in the network.
Therefore it is possible to evaluate a structural compatibility between network elements based on the distance of elements in the module space.
This fact, combined with the observations mentioned in~\secref{sec:char_transp} that 1)~propagation of transmitted signals in cellular networks can be evaluated via local compatibility metrics between network elements and 2)~information on overlapping network modules may serve as a basis for routing, raises the question if, module-based simulated pathways between network elements are correlated to actual pathways observed in the network.

In order to decide, if the observed and simulated pathways are correlated, one may compare observed pathways~$p$, traditional shortest paths~$\hat{p}_{SP}$ and module-based shortest paths~$\hat{p}_{MSP}$: in the latter case the distance~$d_{u,v}$ of 
network elements~$u$ and~$v$ is the Manhattan distance $\sum_k{\left| b_u[k] - b_v[k] \right|}$, where ${b_i}$~is a vector of $M$~components with $b_i[k]$~being the membership strength of network element~$i$~to the module~$k$.

\begin{figure}
\includegraphics[width=\linewidth]{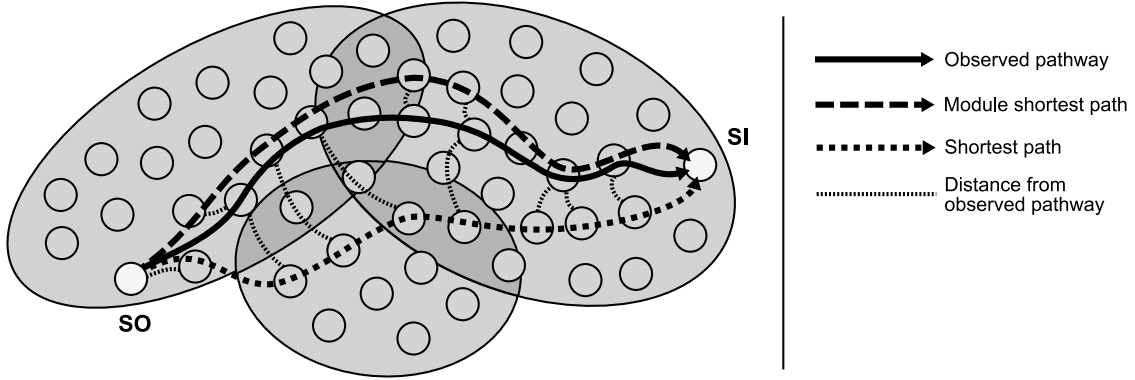}
\caption[Comparing observed, module-based shortest and shortest pathways]{%
An illustrative network with three overlapping modules (marked with ellipses) is shown. Small circles denote network elements, links are not visible.
One observed and two simulated pathways connecting a source element~(SO) with a sink element~(SI) are shown: the actually observed pathway (continuous line), a module-based shortest path with distance metric calculated in module space as described in Section~\ref{sec:sim_pathway} (dashed line) and a traditional shortest path (dotted line). Thin dotted lines indicate the distance between elements
of simulated pathways and the observed pathway. In the scenario of the illustrative figure, the module-based shortest path approximates the observed pathway better than the traditional
shortest path.
}
\label{fig:2}
\end{figure}

To compare an observed pathway~$p$ with a simulated pathway~$\hat{p}$, one may calculate the distance~$d(p, \hat{p})$ by summing the $d(u_{\hat{p}}, p)$ distances between elements~$u_{\hat{p}} \in \hat{p}$ and the pathway $p$,
where $d(u_{\hat{p}}, p)$ is the minimum distance between element~$u_{\hat{p}}$ and any element~$w \in p$. Finally the normalized $d'(p, \hat{p})$ is introduced as
$d(p, \hat{p}) / \left|p\right|$, where $\left|p\right|$~denotes the number of elements in $p$.
Fig.\,\ref{fig:2} shows an illustrative scenario of pathway comparison.

If $d'(p, \hat{p}_{MSP})$ would prove to be generally lower than $d'(p, \hat{p}_{SP})$, then we could conclude that module-based shortest paths are
better approximators of actual network pathways than shortest paths.

Modules often correspond to various functions of the system coded by the network. Therefore, the modular analysis described above may also help us to determine key signaling pathways as inter-modular routes. This becomes especially likely, if we take into account the hierarchical structure of modules, where modules of the original layer are represented as elements of the next layer of hierarchy\cite{msc-palotai-2008,kovacs-2006-pattern}.

\section{Summary}
\label{sec:summary}

We have described that transport processes are significant organizing forces of the network structure and dynamics, and considered
a mechanism of network signaling filtering noise and adapting to newly recognized stimuli.
We investigated the main properties differentiating transport processes, listed expectations towards transport processes and noted that an universally applicable set of optimization criteria does not exist due to criterion-conflicts.
We examined the limits of optimizing network transport for highest possible throughput in the framework of a simple, yet descriptive model of network transport
and described the ways how different network structures could cause, and, interestingly, also relax congestion.

We highlighted the role of overlapping network modules and proposed that exploiting the information on overlapping modules, for example the distance between network elements in the `module space', may help the analysis of routing mechanisms.
Finally, we asked the question if module-based simulated pathways between network elements are correlated to real pathways observed in the network, and
suggested a method for determining the answer.

If modul-based pathways would describe well the real, observed pathways, the identification of key, signaling pathways of complex systems would become possible
using higher layers of the hierarchical modules.
Such knowledge could be utilized in a wide range of fields including traffic control or drug design.

\begin{acknowledgement}
Work in the authors'~laboratory was supported by the EU~(FP6-518230) and the Hungarian National Science Foundation~(OTKA~K69105).
\end{acknowledgement}

\bibliographystyle{adp}

\begin{thebibliography}{[10]}

\bibitem{evolnet-dorogovtsev02}
 \textsc{S.\,N. Dorogovtsev} and  \textsc{J.\,F.\,F. Mendes},
 \jr{Adv. Phys.} \textbf{51}, 1079--1187 (2002).


\bibitem{cellfunc-barabasi04}
 \textsc{A.\,L. Barab\'{a}si} and  \textsc{Z.\,N. Oltvai},
 \jr{Nat. Rev. Genet.} \textbf{5}, 101--113 (2004).


\bibitem{boccaletti-2006}
 \textsc{S.~Boccaletti},  \textsc{V.~Latora},  \textsc{Y.~Moreno},
  \textsc{M.~Chavez},  and  \textsc{D.\,U. Hwang},
 \jr{Phys. Rep.} \textbf{424}, 175--308 (2006).


\othercit
\bibitem{csermely-06}
 \textsc{P.~{Csermely}},
Weak links: {S}tabilizers of complex systems from proteins to social networks
  (Springer, Berlin, 2006).


\bibitem{orginfo-radner93}
 \textsc{R.~Radner},
 \jr{Econometrica} \textbf{61}, 1109--1146 (1993).


\othercit
\bibitem{scott-social-network-analysis}
 \textsc{J.\,P. Scott},
Social Network Analysis: A Handbook (Sage Publications, London, 2000).


\bibitem{fastroad-schutes07}
 \textsc{H.~Bast},  \textsc{S.~Funke},  \textsc{P.~Sanders},  and
  \textsc{D.~Schultes},
 \jr{Science} \textbf{316}, 566 (2007).


\bibitem{Watts:1998}
 \textsc{D.\,J. Watts} and  \textsc{S.\,H. Strogatz},
 \jr{Nature} \textbf{393}, 440--442 (1998).


\bibitem{bionet-zhu07}
 \textsc{X.~Zhu},  \textsc{M.~Gerstein},  and  \textsc{M.~Snyder},
 \jr{Genes Dev.} \textbf{21}, 1010--1024 (2007).


\bibitem{bionet-almaas07}
 \textsc{E.~Almaas},
 \jr{J. Exp. Biol.} \textbf{210}, 1548--1558 (2007).


\bibitem{optimal-structures-cabrales-2002}
 \textsc{R.~Guimer\`a},  \textsc{A.~\`Arenas},  \textsc{A.~D\'iaz-Guilera},
  \textsc{F.~Vega-Redondo},  and  \textsc{A.~Cabrales},
 \jr{Phys. Rev. Lett.} \textbf{89}, 248701 (2002).


\bibitem{cascade-control-motter-2004}
 \textsc{A.\,E. Motter},
 \jr{Phys. Rev. Lett.} \textbf{93}, 098701 (2004).


\bibitem{genanal-gerstein06}
 \textsc{H.~{Yu}} and  \textsc{M.~{Gerstein}},
 \jr{Proc. Natl. Acad. Sci.} \textbf{103}, 14724--14731 (2006).


\bibitem{szalay-2007-581}
 \textsc{M.\,S. Szalay},  \textsc{I.\,A. Kov\'acs},  \textsc{T.~Korcsm\'aros},
  \textsc{C.~B\"ode},  and  \textsc{P.~Csermely},
 \jr{FEBS Lett.} \textbf{581}, 3675--3680 (2007).


\bibitem{palotai-2007}
 \textsc{R.~Palotai},  \textsc{M.\,S. Szalay},  and
  \textsc{P.~Csermely},
 \jr{IUBMB Life} \textbf{60}, 10--18 (2008).


\bibitem{mihalik-2008}
 \textsc{A.~Mihalik},  \textsc{R.~Palotai},  and  \textsc{P.~Csermely},
 \jr{Biochemistry (Hung.)} \textbf{32}, S67 (2008).


\bibitem{csermely-perturb-2009}
 \textsc{M.\,A. Antal},  \textsc{C.~B\"ode},  and  \textsc{P.~Csermely},
 \jr{Curr. Protein. Pept. Sci.} \textbf{10}, 161--172 (2009).


\bibitem{palla-2005-435}
 \textsc{G.~Palla},  \textsc{I.~Der\'enyi},  \textsc{I.~Farkas},  and
  \textsc{T.~Vicsek},
 \jr{Nature} \textbf{435}, 814--818 (2005).


\bibitem{zahavi-75}
 \textsc{A.~Zahavi},
 \jr{J. Theor. Biol.} \textbf{53}, 205--214 (1975).


\bibitem{kovacs-2005}
 \textsc{I.\,A. Kov\'acs},  \textsc{M.\,S. Szalay},  and
  \textsc{P.~Csermely},
 \jr{FEBS Lett.} \textbf{579}, 2254--2260 (2005).


\bibitem{inducedfit-koshland-58}
 \textsc{D.\,E. Koshland},
 \jr{Proc. Natl. Acad. Sci. USA} \textbf{44}, 98--104 (1958).


\bibitem{conf-selection-75}
 \textsc{G.~Careri},  \textsc{P.~Fasella},  and  \textsc{E.~Gratton},
 \jr{CRC Crit. Rev. Biochem.} \textbf{3}, 141--64 (1975).


\othercit
\bibitem{tbaum}
 \textsc{A.\,S. Tanenbaum},
Computer networks: 2nd edition (Prentice-Hall, Upper Saddle River, NJ, USA,
  1988).


\bibitem{wang-survey-2007}
 \textsc{B.\,H. Wang} and  \textsc{T.~Zhou},
 \jr{J. Kor. Phys. Soc.} \textbf{50}, 134--141 (2007).


\bibitem{effrouting-yan-2006}
 \textsc{G.~Yan},  \textsc{T.~Zhou},  \textsc{B.~Hu},  \textsc{Z.\,Q. Fu},  and
   \textsc{B.\,H. Wang},
 \jr{Phys. Rev. E} \textbf{73}, 046108 (2006).


\bibitem{tadic-2006}
 \textsc{B.~Tadic},  \textsc{G.\,J. Rodgers},  and  \textsc{S.~Thurner},
 \jr{Int. J. Bifurcat. Chaos} \textbf{17}, 2363--2385 (2007).


\bibitem{bottlenecks-stanley-2006}
 \textsc{S.~Sreenivasan},  \textsc{R.~Cohen},  \textsc{E.~L{\'o}pez},
  \textsc{Z.~Toroczkai},  and  \textsc{H.\,E. Stanley},
 \jr{Phys. Rev. E} \textbf{75}, 036105 (2007).


\bibitem{respredict-motter-2008}
 \textsc{A.\,E. Motter},  \textsc{N.~Gulbahce},  \textsc{E.~Almaas},  and
  \textsc{A.\,L. Barab\'asi},
 \jr{Mol. Syst. Biol.} \textbf{4}, 168 (2008).


\bibitem{silencing-motter-2008}
 \textsc{T.~Nishikawa},  \textsc{N.~Gulbahce},  and  \textsc{A.\,E.
  Motter},
 \jr{PLoS Comput. Biol.} \textbf{4}, e1000236 (2008).


\bibitem{critsignal-kahn-2006}
 \textsc{C.\,M. Taniguchi},  \textsc{B.~Emanuelli},  and  \textsc{R.\,C.
  Kahn},
 \jr{Nat. Rev. Mol. Cell. Biol.} \textbf{7}, 85--96 (2006).


\bibitem{superhighway-wu-2006}
 \textsc{Z.~Wu},  \textsc{L.\,A. Braunstein},  \textsc{S.~Havlin},  and
  \textsc{H.\,E. Stanley},
 \jr{Phys. Rev. Lett.} \textbf{96}, 148702 (2006).


\bibitem{instabilities-gfeller-2005}
 \textsc{D.~Gfeller},  \textsc{J.\,C. Chappelier},  and
  \textsc{P.~De~Los~Rios},
 \jr{Phys. Rev. E} \textbf{72}, 056135 (2005).


\bibitem{radicchi-2004-101}
 \textsc{F.~Radicchi},  \textsc{C.~Castellano},  \textsc{F.~Cecconi},
  \textsc{V.~Loreto},  and  \textsc{D.~Parisi},
 \jr{Proc. Natl. Acad. Sci. USA} \textbf{101}, 2658--2663 (2004).


\bibitem{impact-community-danon-2008}
 \textsc{L.~Danon},  \textsc{A.~\`Arenas},  and
  \textsc{A.~D\'iaz-Guilera},
 \jr{Phys. Rev. E} \textbf{77}, 036103 (2008).


\bibitem{msc-palotai-2008}
 \textsc{R.~Palotai},
 \jr{MSc. Thesis, Budapest University of Technology and Economics, Budapest,
  Hungary} (2008).


\bibitem{reichardt-2004-93}
 \textsc{J.~Reichardt} and  \textsc{S.~Bornholdt},
 \jr{Phys. Rev. Lett.} \textbf{93}, 218701 (2004).


\bibitem{kovacs-2006-pattern}
 \textsc{I.\,A. Kov\'acs},  \textsc{M.\,S. Szalay},  \textsc{P.~Csermely},  and
   \textsc{T.~Korcsm\'aros},
 \jr{{P}atent application number: WO2007093960} (2006).


\bibitem{nepusz-2007}
 \textsc{T.~Nepusz},  \textsc{A.~Petr\'oczi},  \textsc{L.~N\'egyessy},  and
  \textsc{F.~Bazs\'o},
 \jr{Phys. Rev. E} \textbf{77}, 016107 (2007).


\bibitem{zhang-2007}
 \textsc{S.~{Zhang}},  \textsc{R.\,S. {Wang}},  and  \textsc{X.\,S.
  {Zhang}},
 \jr{Phys. A} \textbf{374}, 483--490 (2007).


\end{thebibliography}


\newcommand{\noopsort}[1]{}
\providecommand{\WileyBibTextsc}{}
\let\textsc\WileyBibTextsc
\providecommand{\othercit}{}
\providecommand{\jr}[1]{#1}
\providecommand{\etal}{~et~al.}

\end{document}